\documentclass[usenatbib]{mn2e} 

\usepackage{caption}
\usepackage{tabularx}
\usepackage{lmodern}
\usepackage{color}
\usepackage{graphicx}
\usepackage{txfonts}
\usepackage{multirow}
\usepackage{supertabular}
\usepackage{longtable}
\usepackage{rotating}
\usepackage{rotating} 
\usepackage{lscape} 
\usepackage{amssymb}
\usepackage{dcolumn}

\usepackage{arydshln}
\usepackage{natbib}

\DeclareMathAlphabet{\mathsc}{OT1}{cmr}{m}{sc}
\def\testbx{bx}%
\DeclareRobustCommand{\ion}[2]{%
\relax\ifmmode
\ifx\testbx\f@series
{\mathbf{#1\,\mathsc{#2}}}\else
{\mathrm{#1\,\mathsc{#2}}}\fi
\else\textup{#1\,{\mdseries\textsc{#2}}}%
\fi}

\def\h2{$\rm H_2$}

\def\sys{J1619$+$3342}

\def\lyb{\ensuremath{{\rm Ly}\beta}}
\def\kms{km\,s$^{-1}$}

\def\zabs{$z_{\rm abs}$}
\def\zem{$z_{\rm em}$~}

\def\hi{H~{\sc i}}

\begin{document}

\title[Nearest \h2 absorber] {Molecular hydrogen from $z=0.0963$ DLA Towards the QSO \sys \thanks{Based on observations made with the NASA/ESA Hubble Space Telescope,
    obtained from the data archive at the Space Telescope Science Institute, which
    is operated by the Association of Universities for Research in Astronomy, Inc.,
    under NASA contract NAS 5-26555.}}
\author[Srianand et. al.]{R. Srianand$^{1}$\thanks{email:anand@iucaa.ernet.in},H. Rahmani$^{1,2}$, S. Muzahid$^{3}$ and V. Mohan$^{1}$\\
$^{1}$ Inter-University Centre for Astronomy and Astrophysics, Post Bag 4,  Ganeshkhind, Pune 411\,007, India \\
$^{2}$ School of Astronomy, Institute for Research in Fundamental Sciences (IPM), PO Box 19395-5531, Tehran, Iran\\
$^{3}$ The Pennsylvania State University, 413 Davey Lab, University Park, State College, PA 16802, USA \\
}
\pagerange{\pageref{firstpage}--\pageref{lastpage}} \pubyear{2012}
\maketitle
\label{firstpage}

\begin {abstract}
{
We report the detection of \h2\  in a \zabs = 0.0963 Damped Lyman-$\alpha$ 
(DLA) system towards \zem = 0.4716 QSO J1619+3342. This DLA has 
log~$N$(H~{\sc i}) = 20.55$\pm$0.10, 18.13$\le$log~$N$(H$_2$)$\le$18.40,  
[S/H] = $-0.62\pm0.13$, [Fe/S] = $-1.00\pm0.17$ and the molecular 
fraction $-2.11\le log [f({\rm H_2})]\le -1.85$. The inferred gas kinetic temperature
using the rotational level population is in the range 95$-$132 K.
We do not detect C~{\sc i} or C~{\sc ii$^*$} 
absorption from this system. Using R and V band deep images we identify a sub-L$_*$ galaxy
at an impact parameter of 14 kpc from the line of sight, having consistent
photometric redshift, as a possible host for the absorber.
We use the photoionization code {\sc cloudy} to get the
physical conditions in the \h2 component using the observational constrains
from \h2, C~{\sc i}, C~{\sc ii$^*$} and Mg~{\sc i}. All the observations
can be consistently explained if one or more of the following is true:
(i) Carbon is underabundant by more than 0.6 dex as seen in halo stars with Z$\sim$0.1 Z$_\odot$, 
(ii) H~{\sc i} associated with \h2 component is less than 50\% of the H~{\sc i} measured along the
line of sight and (iii) the \h2 formation rate on the dust grains is
at least a factor two higher than what is typically used in analytic calculations
for Milky Way interstellar medium. Even when these are satistifed, the gas
kinetic temperature in the models are much lower than what is inferred from
the ortho-to-para ratio of the molecular hydrogen. Alternatively the high 
kinetic temperature could be a consequence of contribution to the gas heating
from non-radiative heating processes seen in hydrodynamical simulations.
}
\end{abstract}

\section{Introduction}
Damped Lyman-$\alpha$ systems (DLAs) are the highest H~{\sc i} column density 
absorbers seen in QSO spectra, with $N$(H~{\sc i})$\ge$2$\times$10$^{20}$\,cm$^{-2}$
\citep[see][for a review]{Wolfe05}.  
These absorbers trace bulk of the neutral hydrogen at 2$<z<$3 
\citep{Prochaska05,Noterdaeme09dla,Noterdaeme12dla} and have been conjuctured to
be originating from the gas associated with high-$z$ galaxies and protogalaxies.
The link between DLAs and galaxies can be established by directly 
detecting galaxies and/or showing that the prevailing physical conditions
in the absorbing gas are consistent with those seen in a typical galactic
Interstellar Medium (ISM).

Our understanding of the physical conditions in DLAs is primarily based on optical 
absorption-line studies, involving the detection of low-ionization metal transitions and, 
in a few cases, \h2, HD and CO molecular absorption \citep[see for example,][]{Ledoux03, Srianand08,Varshalovich01,Noterdaeme08,Noterdaeme08hd,Noterdaeme09co}. 
DLA subcomponents in which H$_2$ absorption is detected, 
have typical temperature and density of 153$\pm$78 K and $n_{\rm H}$ = 10$-$200\,cm$^{-3}$,
 respectively \citep[][]{Srianand05}. 
 The gas producing \h2 absorption probably traces diffuse molecular gas
that is compact \citep{Balashev11} containing only
a small fraction of H~{\sc i} measured using the DLA profile \citep{Srianand12}. 
When CO is detected one gets a chance to probe high-$z$ translucent regions \citep{Noterdaeme10co} 
and rotational excitations of CO are mainly dominated by the pumping by cosmic microwave background \citep{Srianand08,Noterdaeme11cmb}. The 
C~{\sc ii$^*$} absorption, detected in roughly 50\% of the high-$z$ DLAs, is also used to infer 
physical conditions \citep{Wolfe08}. 
The inferred radiation field, using some of these indicators when present, is similar to the mean Galactic field probably originating from local star formation 
activities.
However, direct detection of host galaxies of these high-$z$ DLAs is needed to firmly
establish a direct link between galaxy properties and the physical conditions inferred
using various indicators discussed here.

Despite several attempts a firm link between  galaxies and DLAs 
is not established at high-$z$, mainly due to the paucity of direct detection of
galaxies. Only few DLAs with high metallicity and high $N$(H~{\sc i}) are detected
through the Lyman-$\alpha$ emission \citep[see][and references there for the summary]
{Fynbo13}. Observed host galaxies are usually found at large impact parameters
suggesting that the absorbing gas is located well outside the luminous part of the galaxy
and large-scale winds and outflows are playing a vital role in populating the 
circumgalactic regions of these high-$z$ galaxies with cold gas \citep[]{Bouche13, Fynbo13, Kashikawa14}.
Such a picture is also 
supported by simulations \citep[see][]{Pontzen08,Altay11}.
However, host galaxies are not detected for most of the DLAs that show \h2 and/or C~{\sc ii$^*$}
at high-$z$.

At low-$z$ (i.e., $z\le 1$) it is relatively easy to identify the galaxy counterparts of 
DLAs and sub-DLAs \citep[][]{Burbidge96, Lebrun97, Rao03, Meiring11,Battisti12}. 
Associated galaxies span a wide range of morphology and impact parameters. Establishing
the connection between physical conditions derived from above mentioned indicators and
the star formation activities in low-$z$ galaxies will enable us to interpret the
high-$z$ DLA observations. 
However, only a few low-$z$ DLAs are known \citep[see for example,][]{Rao06} and only two \h2 detections are known at $z\le1$ \citep[][]{Crighton13, Oliveira14}. Thanks to HST/COS the situation is improving rapidly \citep[][]{Meiring11,Battisti12}. 
Here we present the detection of \h2 in a \zabs = 0.0963 DLA towards the QSO J1619+3342 and study 
the physical conditions prevailing in this system.

\section{Observations and data reduction}      

\begin{table*}
\caption{Results of single cloud curve of growth for \h2 absorption}
\begin{tabular}{lccccccccc}
\hline
\hline
Level & Transition & $\lambda_r$ & $f$ & $\lambda_r$(range) & $W_0$&\multicolumn{2}{c}{$\delta W_0$}  & \multicolumn{2}{c}{log[N(cm $^{-2}$)]}\\
      &         &            &      &                   &       &  (sta)          &  (sys)            & COG$^+$ & VPFIT\\
      &         &  (\AA)     & ($10^{-2}$)     & (\AA)             & (\AA) & (\AA)           &(\AA)              &     &   \\
\hline
J = 0 & ${\rm L_1R_0}$ & 1092.1952 & 0.578 & 1092.11-1092.50 & 0.157 & 0.008 & 0.008 & 17.54$-$17.90 &18.17$\pm$0.04    \\
      & ${\rm L_2R_0}$ & 1077.1387 & 1.170 & 1077.06-1077.33 & 0.147 & 0.006 & 0.008 &             &                  \\
      & ${\rm L_4R_0}$ & 1049.3673 & 2.310 & 1049.23-1049.54 & 0.172 & 0.011 & 0.008 &             &                  \\
J = 1 & ${\rm L_1R_1}$ & 1092.7323 & 0.378 & 1092.65-1092.94 & 0.123 & 0.006 & 0.012 & 18.00$-$18.22 & 18.36$\pm$0.04   \\
      & ${\rm L_2P_1}$ & 1078.9254 & 0.392 & 1078.81-1079.14 & 0.164 & 0.008 & 0.012 &             &                  \\
      & ${\rm L_3R_1}$ & 1063.4601 & 1.190 & 1063.35-1063.70 & 0.189 & 0.007 & 0.012 &             &                  \\
      & ${\rm L_4R_1}$ & 1049.9597 & 1.550 & 1049.84-1050.16 & 0.193 & 0.010 & 0.009 &             &                  \\
      & ${\rm L_5P_1}$ & 1038.1570 & 0.866 & 1038.00-1038.30 & 0.171 & 0.012 & 0.010 &             &                  \\ 
J = 2 & ${\rm L_0P_2}$ & 1112.4959 & 0.069 & 1112.37-1112.70 & 0.048 & 0.007 & 0.025 & 15.44$-$16.88 & 15.97$\pm$0.25   \\
      & ${\rm L_1P_2}$ & 1096.4383 & 0.236 & 1096.34-1096.60 & 0.070 & 0.007 & 0.013 &             &                  \\
      & ${\rm L_2P_2}$ & 1081.2659 & 0.469 & 1081.21-1081.42 & 0.057 & 0.009 & 0.010 &             &                  \\
      & ${\rm L_2R_2}$ & 1079.2254 & 0.681 & 1079.14-1079.36 & 0.075 & 0.007 & 0.010 &             &                  \\
      & ${\rm L_3P_2}$ & 1066.9006 & 0.709 & 1066.84-1067.05 & 0.083 & 0.007 & 0.010 &             &                  \\
      & ${\rm L_3R_2}$ & 1064.9947 & 1.060 & 1064.92-1065.19 & 0.083 & 0.008 & 0.013 &             &                  \\
      & ${\rm L_4P_2}$ & 1053.2842 & 0.902 & 1053.14-1053.44 & 0.094 & 0.010 & 0.015 &             &                  \\
      & ${\rm L_5P_2}$ & 1040.3672 & 1.020 & 1040.20-1040.47 & 0.095 & 0.013 & 0.012 &             &                  \\
      & ${\rm L_5R_2}$ & 1038.6901 & 1.660 & 1038.59-1038.77 & 0.072 & 0.011 & 0.008 &\\
J=3   & ${\rm L_4R_3}$ & 1053.9760 & 1.340 & 1053.80-1054.15 & 0.072 & 0.010 & 0.019 & 14.70$-$15.69 & 15.27$\pm$0.29 \\
      & ${\rm L_5R_3}$ & 1041.1588 & 1.580 & 1041.05-1041.22 & 0.048 & 0.010 & 0.010\\

 \hline
\end{tabular}
\flushleft{$^+$ 1 $\sigma$ range in column density obtained using 1$\sigma$ range in $b$ and $ W_0$.}
\label{tab1}
\end{table*}

The ultraviolet (UV) spectrum of the QSO SDSS J161916.54+334238.4 
($z_{em}\sim0.4716$; refer to as J1619+3342 in this paper) was obtained using 
the Cosmic Origins Spectrograph (COS) on board the 
{\it Hubble Space Telescope} ($HST$) during observing cycle-17, 
under program ID: 11598 (PI: Jason Tumlinson). The observations consist 
of G130M (5.3 ks) and G160M (8.8~ks) far-UV (FUV) grating integrations 
at a medium resolution of $R \sim 18,000$ (FWHM $\sim$ 18 km s$^{-1}$). 
The data were retrieved from the $HST$ archive and reduced using the 
STScI {\sc calcos} (v2.17.3) pipeline software. The individual reduced 
$x1d$ files were flux calibrated. {The alignment and co-addition of the 
separate G130M and G160M exposures were done using the software developed 
by \citet{Danforth10}\footnote{http://casa.colorado.edu/$\sim$danforth/science/cos/costools.html}}. 
The exposures were weighted by the integration time while co-adding 
the flux calibrated data. The final coadded spectrum covers  the wavelength range 
1134 -- 1796 \AA\ with the signal-to-noise ratio $S/N \sim$ 8--12 per 
resolution element. Each COS resolution element is sampled by six 
raw pixels. We therefore binned the spectrum by 3 pixels and perform 
most of our analysis/measurements using this binned data. Continuum normalization 
was done by fitting the line free regions with a smooth lower order polynomial. 

The COS wavelength calibration known to have uncertainties at the level of 10--15 \kms. 
Regions of spectrum that are recorded near the edges of the detector segment are more prone 
to have such erroneous wavelength solution \citep[]{Savage11,Meiring13}. It is also known
that the line spread function (LSF) of the COS spectrograph is not a Gaussian. A 
characterization of the non-Gaussian COS LSF is found in \citet{Ghavamian09} and 
subsequently updated by \citet{Kriss11}. We adopt the \citet{Kriss11} LSF for our Voigt 
profile fitting analysis. Interpolated LSF at the line center were convolved with the model Voigt 
profile while fitting an absorption line using the 
{\sc vpfit}\footnote{http://www.ast.cam.ac.uk/$\sim$rfc/vpfit.html} code.   

We used the 2-m telescope at IUCAA Girawali Observatory (IGO) to image
the field in R and V bands using IUCAA Faint Object Spectrograph
and Camera (IFOSC) on 20-21 March 2013. Total exposure times
are 3900s and 4900s in V and R band respectively. A typical
seeing during these observations was in the range 1.3 to 1.4 arcsec. 
However, as these images are much deeper than those of SDSS, we  use them
to identify nearby galaxies to the QSO line of sight.

\section{Properties of the DLA from the absorption line analysis}

\subsection{Measurements based on atomic lines}

In this sub-section we derive physical conditions using the metal line absorption
originating from neutral and singly ionized species.

\subsubsection{Metallicity and dust depletion}

Column density and metallicity measurements for this DLA are presented in detail 
in Table~3 of \citet{Battisti12}. These authors find log~$N$(H~{\sc i}) = 20.55$\pm$0.10, 
[S/H] = $-0.62\pm0.13$, [N/H] = $-1.74\pm0.16$\footnote{Only N~{\sc i} column density is used to derive [N/H] and we 
do not include contribution from N~{\sc ii}.}, [P/H] = $-0.81\pm0.21$ and [Fe/S] = $-1.00\pm0.17$.  In addition to these,
we measure log~$N$(Ar~{\sc i}) = $14.12\pm0.11$ and obtain 3$\sigma$ upper limits\footnote{ Using Ar~{\sc i}$\lambda\lambda 1048,1066$, Cl~{\sc i}$\lambda 1347$, C~{\sc i}$\lambda1650$ and C~{\sc ii}$^*\lambda1335$
transitions.}: 
log~$N$(Cl~{\sc i})$\le$13.09, log~$N$(C~{\sc i})$\le$13.30 and log~$N$(C~{\sc ii}$^*$)$\le$13.12 {\citep{Battisti12}} 
using the COS spectrum. Based on Ar~{\sc i}, which is the dominant ionization stage of Ar in the
neutral gas, we get [Ar/H] = $-0.83\pm0.16$. Note [Ar/S] = $-0.21\pm0.14$ also confirms
that the Ar is mostly in Ar~{\sc i} if we assume that the intrinsic [Ar/S] is solar. 
This suggests that the gas may not be ionized by a 
hard radiation field \citep[][]{Vladilo03}. From \citet{Battisti12} we also have 
log~$N$(Mg~{\sc i}) = $12.40\pm0.14$, log~$N$(Ca~{\sc ii}) = 12.42$\pm$0.02 and
log~$N$(Ti~{\sc ii})=11.90$\pm$0.04 based on their KECK/HIRES spectrum that
also shows the metal absorption to be present in two distinct velocity components
separated by $\sim 9$ \kms. Note at the HST/COS's resolution the two components
are not resolved. Therefore, we use the total column density for all the analysis presented
below.  The measured [Ti/S] = $-0.90\pm0.10$ suggests that, for the 
measured $N$(H~{\sc i}), the Ti depletion is much less than what is typically 
seen in the Milky Way disk and in the lower end of the measurements towards Milky Way 
outer halo or Large Magellanic clouds \citep[LMC,see Fig 3 of,][]{Welty10}.
They also suggested that the sight lines with such values of [Ti/S]
may be related to gas having low density and low molecular
fraction (i.e., $f({\rm H_2}) = {2N({\rm H_2})/[N(\ion{H}{I})+2N({\rm H_2})]} \le$ 0.1).

It is found that the ratio of $N$(Ti~{\sc ii}) to $N$(Ca~{\sc ii}) remains nearly
constant in the local measurements. This ratio has a value
of $\sim0.3$ in the Galactic ISM and $\sim 0.9$ in the case of LMC and Small Magellanic Clouds 
\citep[SMC, see][]{Cox06,Cox07}. 
It is
interesting to note that in the present case we find this ratio to be 0.3 consistent
with the Milky Way measurements. 
We also notice that the ratio of $N$(Fe~{\sc ii}) to $N$(Ti~{\sc ii}) is close to
solar suggesting similar depletion for Fe and Ti and the absence of major ionization
effects. In summary, purely based on metallicity measurements we can conclude
that the average metallicity [i.e., [S/H]$\sim-0.62$] is similar to the mean value measured for SMC sight lines.
The depletion pattern is intermediate between lower Milky way halo gas and LMC sight lines. 
If the strong correlation seen between average line of sight density and Ti depletion
is applicable then the gas density in the present DLA may be lower than that 
typically seen in the Milky Way disk \citep[See Fig 2 of][]{Welty10}.

All the above quoted abundance measurements are based on dominant neutral or singly ionized species without 
applying ionization corrections. However, discussions presented above suggest that the ratios
of singly ionized species are not severely affected by ionization effects.
Assuming intrinsic [Fe/S] to be solar we get the column density of dust in Fe, log~$N$(Fe)$_{\rm dust}$ = 14.83
and dust to gas ratio log~$\kappa = -0.74$. At high-$z$, DLAs with such high values of metallicity, $\kappa$ and
$N$(Fe)$_{\rm dust}$  tend to have detectable \h2 \citep{Ledoux03,Petitjean06,Noterdaeme08}. 

\subsubsection{The absence of C~{\sc ii}$^*$ fine-structure line:}

In the Galactic ISM, [C~{\sc ii}] 158$\mu$ line emission is a dominant coolant in the neutral
gas. If this is also the case in high-$z$ DLAs then one can estimate the gas cooling
rate per hydrogen atom ($l_c$) using $N$(C~{\sc ii}$^*$). Under steady state conditions, one
can use this to infer the gas heating rate and hence the insitu starformation rate 
\citep[][]{Wolfe03a}. 
Using the distribution of $l_c$ in high-$z$ DLAs \citet{Wolfe08} have suggested the existence
of two populations of C~{\sc ii}$^*$ absorbers which they called ``low-'' and ``high-cool'' 
absorbers.
In the present case we do not detect the  C~{\sc ii}$^*$ absorption. 
The 3$\sigma$ upper limit on $N$(C~{\sc ii}$^*$) corresponds to an upper limit of
log~${l_c} \le -27.03$\footnote{$l_c$ is defined
in units of erg s$^{-1}$ per hydrogen atom}. This is much lower
than what is typically seen in the Milky Way disk and low, intermediate and high
velocity clouds \citep{Lehner04}. The upper limit also puts the present DLA amongst the ``low-cool''
population identified by \citet{Wolfe08}.  \citet{Wolfe08} have suggested that the 
``low-cool'' population may either related to (i) Warm neutral medium (with T$\sim$ 8000 K) in 
ionization equilibrium with the meta-galactic radiation field or (ii) metal poor compact region
with continuously on-going in situ starformation. However, what is interesting to note is that
the measured [S/H] in the present case is much higher than that typically seen in high-$z$
``high-'' (mean [$\alpha$/H] = $-1.06\pm0.15$)  and ``low-cool'' (mean [$\alpha$/H] = $-1.74\pm0.19$) 
absorbers. Therefore, if metallicity is the crucial factor that discriminates this two population
the present DLA should belong to the higher end of the ``high-cool'' DLAs.

Next we look at the relationship between $l_c$ and \h2 detections at high-$z$.
At high-$z$, C~{\sc ii}$^*$ is detected almost in all 
\h2 systems (with lower $N$(C~{\sc ii}$^*$) compared to the present limit in
few cases) whenever the expected wavelength range is not contaminated by absorption lines from other intervening
gas. However, we note that about 36\% of the \h2 DLAs have  log~${l_c} \le -27.03$. 
This suggests that even among \h2 detected DLAs (with signatures of low temperatures in the
\h2 components) low $l_c$ values as seen in the present case are not that uncommon. 

Direct measurement of $N$(C~{\sc ii$^*$})/$N$(C~{\sc ii}) is important to draw physical conditions in
the gas.
Unfortunately, it is difficult to measure the C~{\sc ii} column density directly from the
COS data. Firstly the C~{\sc ii}$\lambda$1334 line is saturated and partially blended with \lyb\
line from \zabs = 0.4268. Secondly the absorption component has a velocity shift of about
10 \kms\ with respect to the \zabs\ defined by other metal lines.  Based on the velocity
off-set between the Lyman series lines of \zabs = 0.4268 we confirm that C~{\sc ii}$\lambda$1334
is highly saturated but not showing damping wings. The C~{\sc ii}$\lambda$1036 line is 
covered by the COS spectrum. Unfortunately this line is blended with another strong absorption
line.  Therefore, direct estimation of C~{\sc ii} column density is difficult.
So we use indirect estimations of  $N$(C~{\sc ii}) for discussions presented below.

If we assume that 
the abundance ratio [C/S] is close to solar and no depletion of C into dust grains then we 
expect log~$N$(C~{\sc ii}) = 16.39. This gives log~$N$(C~{\sc ii}$^*$)/$N$(C~{\sc ii}) $\le -3.27$.
 We wish to note here that this inferred column density is
allowed by the observed profile if we apply a necessary velocity shift to match the
\lyb\ profile of \zabs = 0.4268 system with rest of the Lyman series lines. 
However, if the C enhancement in this DLA follows what one sees in intermediate and low metallicity
halo stars in our Galaxy then we expect [C/O]$\sim-0.7$ for the metallicity observed in the
present DLA \citep[][]{Akerman04,Fabbian09,Dutta14}. In such a case we will have 
log~$N$(C~{\sc ii}$^*$)/$N$(C~{\sc ii}) $\le -2.57$. Note any depletion of C with respect to S
due to dust depletion will further increase this ratio up to 0.4 dex \citep[see for example][]{Sofia11}.

As can be seen from the discussions given in \citet{Srianand05}, the predicted
value of log~$N$(C~{\sc ii}$^*$)/$N$(C~{\sc ii})  is in the range $-$2.62 to $-$2.20 in the case of
a typical Cold Neutral Medium (CNM)  and in the range $-$2.65 to $-$3.17 in the case of a typical
Warm Neutral Medium (WNM) of our Galaxy.  However, for the typical abundance and dust depletion prevailing
in high-$z$ DLAs the corresponding ranges are $-$2.26 to $-$2.08 for CNM and $-$3.39 to $-$2.70
for WNM.
It is clear that if the [C/S] in this DLA is close to solar then the absence of C~{\sc ii}$^*$
is consistent with the WNM solution. However, if [C/S] is close to intermediate metallicity
stars then the measured C~{\sc ii}$^*$ upper limit will allow for a CNM solution.
We come back to this issue later when we discuss photo-ionization models for this system.

\subsubsection{The absence of C~{\sc i} absorption line:}

\citet{Srianand05} have suggested a possible connection between the presence of C~{\sc i} absorption
and \h2 detection. Among \h2 detections available till date \citep[excluding the CO systems 
detected through the presence of strong C~{\sc i} lines in SDSS spectrum, see][]{Srianand08, Noterdaeme10co,Noterdaeme11cmb}
we notice that about 73\% of the high-$z$ \h2 systems have detectable C~{\sc i} lines. Note most of the
detected C~{\sc i} lines have column densities much lower than the upper limit of C~{\sc i} we find
for the present DLA. Therefore, absence of C~{\sc i} absorption in the present system 
is not that uncommon even in high-$z$ DLAs with \h2 detections . Following \citet{Srianand05} we can write,
\begin{equation}
{n_e\over \Gamma} = 4.35\times 10^{11}~{N({\rm \ion{C}{i}})\over N({\rm \ion{C}{ii}})}~\bigg{(} {T\over 10^4}\bigg{)}^{0.64}.
\end{equation}
We can use the observed value of $N$(C~{\sc i})/$N$(C~{\sc ii}) to place a constraint
of $n_e/\Gamma$, where $\Gamma$ and $n_e$ are the C$^0$ photoionization rate and electron density
respectively. When we assume $T= 100$ K [as inferred in the next session],
we get  $n_e/\Gamma\le 1.8\times 10^7$ s cm$^{-3}$ and  $n_e/\Gamma \le 9.0\times 10^7$ s cm$^{-3}$
respectively for $N$(C~{\sc ii}) obtained using [S/C] close to solar and [S/C] close to what is seen
in intermediate metallicity halo stars. When we use the mean $n_e$ measured by \citet{Welty03}
and $\Gamma$ from \citet{Pequignot86} we get the mean $n_e/\Gamma \sim 7 \times 10^8$ s cm$^{-3}$
for the diffuse interstellar medium in Milky Way. Therefore, either the absorber has larger ionization rate or has lower $n_e$ compared to what is typically seen in CNM of our Galaxy. However, if we
use $\Gamma = 2\times10^{-10}$ s$^{-1}$ (a typical value for the Milky Way ISM) 
then we get $n_e$ $\le$ 0.004 and 0.012 cm$^{-3}$
respectively for the two inferred values of $N$(C~{\sc ii}). These are
consistent with what one sees in 40\% of the components in high-$z$ DLAs with
C~{\sc i} detections \citep[See Table 4 of][]{Srianand05}.

\subsubsection{High ions and velocity structure in the absorbing gas:}

\citet{Battisti12} have reported the detection of N~{\sc ii}, Si~{\sc iii} and Fe~{\sc iii} 
in their
HST/COS spectrum. From their Fig.~10, it is apparent that there is no clear velocity 
segregation between low ion and Fe~{\sc iii}. We also notice that the 
N~{\sc ii} absorption profile very much follows that of the singly ionized species. 
This means that the gas we are dealing with is a multiphase medium and some fraction 
of the observed column density of singly ionized
species may originate from N~{\sc ii} or F~{\sc iii} phase.  Weak Si~{\sc iv} and C~{\sc iv}
absorption are also detected in this component. A strong C~{\sc iv} absorption is detected
at -100 \kms with respect to the main component. No other ion is detected in this component.
As in high-$z$ DLAs this component may probably be related to outflows \citep{Fox07a} and may not contain
much of $N$(H~{\sc i}) measured from the DLA profile.  
As discussed before consistent abundance ratios inferred from singly ionized species are
a good indication for the absence of strong influence from ionization effects in this phase.
However, the lack of information on
H~{\sc i} column density of this phase will hamper our attempt to model the system
using photoionization equilibrium.

\subsection{\h2 measurements:}
\begin{figure}
\centerline{
\includegraphics[width=8cm,bb=25 120 600 710,clip=,angle=0]{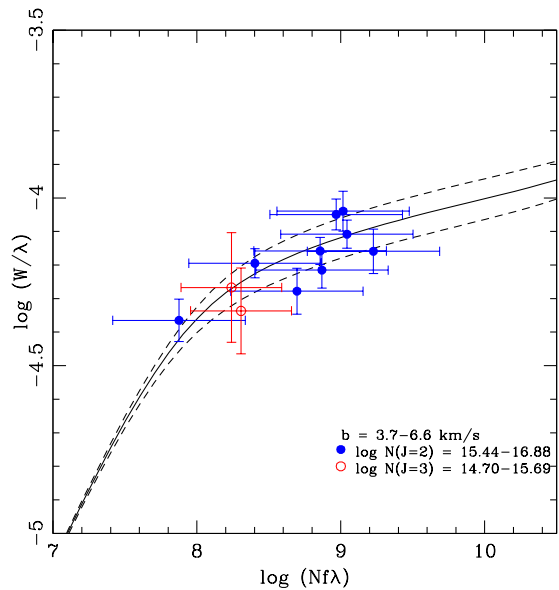} 
}
\caption{Results of single component curve of growth analysis. The
best fitted curve for the absorption from $J=2$ and $J=3$ levels are shown 
overlayed with data points. The ordinate
errors are based on the measured equivalent width errors and the
abscissa errors are due to column density errors from our COG analysis.}
\label{fig_cog}
\end{figure}

Molecular hydrogen is detected in the DLA in four rotational levels (i.e., up to $J\le 3$).
We have identified a set of uncontaminated \h2 lines for the column density measurements.
Details of these \h2 lines are summarized in Table~\ref{tab1}. In this
table the first 4 columns give the rotational level, species ID, rest wavelength and
oscillator strength respectively. As one expects the b-parameter of the \h2 components 
to be smaller than the spectral resolution of COS one has to be careful about the hidden 
saturation effects while measuring column densities. In addition, the apparent shifts 
between different absorption lines due to wavelength calibration uncertainties and non-gaussian
LSF, if not 
taken into account properly, can lead to wrong column density estimates when one uses Voigt profile
fitting codes. So we use both curve of growth (COG) and VPFIT to constrain the b parameter
and measure the \h2 column densities.

\subsubsection{Curve of growth analysis:}

In Table~\ref{tab1}, we summarise rest equivalent width ($W_0$) measurements for the clean lines
and associated errors ($\delta W_0$). We estimate two sets of
errors for the observed equivalent widths. These are statistical error
due to uncertainties in the flux measurements ($\delta W_0 (sta)$) and
systematic error from the continuum placement uncertainties ($\delta W_0 (sys)$).
We measure the latter by using different continuum normalisations. For the
COG analysis discussed below we use combined errors.

The rest wavelength range used to get the
equivalent width is also given in the 5th column of Table~\ref{tab1}.
It is clear from this table that equivalent widths from a given $J$
levels are not scaling  as $\lambda f$ as expected for the lines in the
linear portion of the curve of growth. This suggests that saturation effects
are important. The $J=2$ level has several
clean transitions observed so that one can get good constrints on $N$ and
$b$ using the single cloud curve of growth.

When we use  measured rest equivalent widths and combined errors of all 
9 transitions observed 
for the $J=2$ level a best fit is obtained for 
log~$N$(\h2,$J=2$) = 16.16$\pm$0.72 and $b$ = (5.2$\pm$1.4) \kms 
$\chi^2_\nu$ = 0.69 (see Fig.~\ref{fig_cog}). 
The derived $b$-parameter is of the order of one third of the spectral resolution of HST-COS.
For simplicity we assume this $b$-parameter range for all the $J$ levels 
to measure the column densities. In general this need not be the 
case as b may depend on the $J$ levels \citep[see for example,][]{Noterdaeme07}.
For $b$ value similar to  or less that what we obtained above the $J=0$ and
$J=1$ lines are in the damped part of the curve of growth. Using the strongest
two transitions for $J=0$ and $J=1$ and assuming these lines are in the damped part of the COG, 
we estimate
the range in column densities for these levels. 
The column density range for each $J$ resulting from this analysis
are also summarized in Table~\ref{tab1}.

\subsubsection{Voigt profile analysis}
\begin{figure*} 
\centering
\begin{tabular}{cc}
\includegraphics[width=15cm,height=13cm,bb=56 10 774 591,clip=,angle=0]{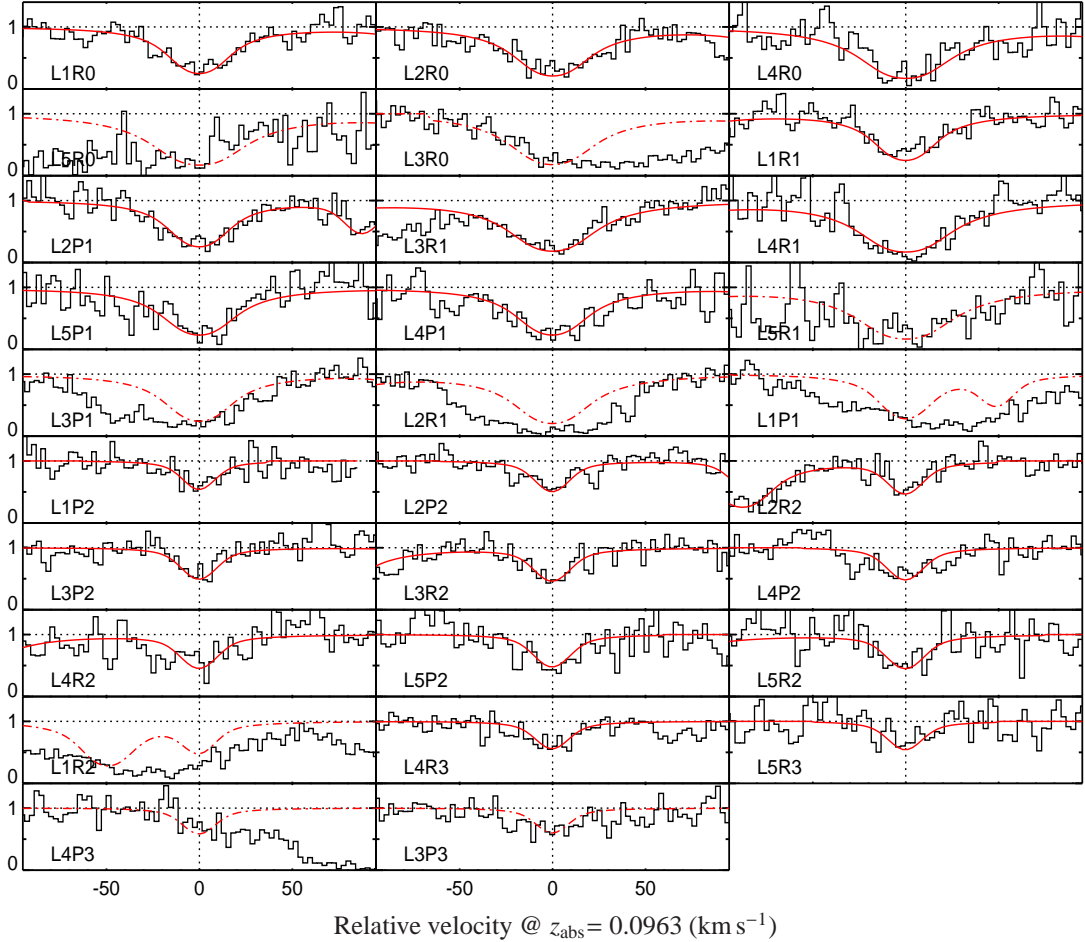} 
\end{tabular}
\caption{Single component voigt profile fits to \h2 absorption detected
in the \zabs = 0.0963 DLA towards J1619+3342. The solid smooth profile over-plotted on the 
observed data is our best fit model. The dashed profiles are the model
predictions in cases where the \h2 transition is not used in the fit 
as they are blended with absorption from other systems.}
\vskip -14.0cm
\begin{picture}(400,400)(0,0)
\put( 120, 52){\large Relative velocity @ \zabs = 0.0963 (\kms)}
\end{picture}
\label{fig_h2}
\end{figure*}

We fitted the same set of clean \h2 lines with a single component
voigt profile using the {\sc vpfit} code. In these fits we tried keeping same
$b$-parameter for absorption from all $J$ levels. We constrained the
column density to be the same for a given $J$ level and allow it to be
different for different $J$ levels. As wavelength scales are not
accurate we did allow the redshifts of individual transitions to be
different.  Best fit is obtained for $b$ = 4.1$\pm$0.4 \kms with
a reduced $\chi^2 = 0.63$. The best fitted column densities from
the voigt profile fits are summarized in the last column of 
Table~\ref{tab1}.  The best fitted profile overlayed on the observed
data are shown in Fig.~\ref{fig_h2}. For consistency check, in this
figure we also show the predicted profiles of the blended \h2 lines
with the dotted profiles.

The best fitted $b$ parameter
is slightly lower and $N$($J$) values are slightly higher than 
the ones obtained using COG analysis. The column density errors are small in the case
of measurements using {\sc vpfit}. This is mainly because the 
quoted errors are only statistical and systematic errors from
the continuum placement uncertainties are not included.
We also consider voigt profiles fits allowing the $b$ parameter to
be different for different $J$ levels. We find the column
densities derived for $J=0$ and 1 levels are consistent with
what we quote in Table~\ref{tab1} even though a lower
value of $b$ parameter is preferred. This is the indication that some of these lines
are in the damped part of the COG. The value of $b$ is 
ill-constrained in the case of $J=3$ as we have only two clean
lines used for the fit.  

\subsubsection{Molecular fraction \& Kinetic temperature}

{Here we estimate physical parameters using \h2 column densities obtained
with COG analysis that propertly takes care of the continuum placement uncertainties.}
The total \h2 column density from the COG analysis is log $N$(\h2) = 18.13$-$18.40
and 
the molecular fraction, $-2.11\le$log~$f$(\h2)$\le-1.85$\footnote{The $f$(\h2) measured
using voigt profile analysis is 0.2 dex higer than that obtained using COG. However, the
excitation temperature derived using both the methods are consistent.}.
In the Galactic disk log~$f$(\h2) is expected to be $\le -4$ for the $N$(\hi) observed in this
system \citep{Savage77}. 
Such high values of $f$(\h2) are not seen for log~$N$(\hi) = 20.55, even
in the case of Magellanic clouds \citep[See Fig~8 of][]{Tumlinson02}.
The observed high value of log~$f$(\h2) is consistent
with what is seen in the high lattitude (i.e., $|b|\ge20^\circ$) 
clouds in the halo of our Galaxy \citep[see Fig~6 of][]{Gillmon06} and also 
in a couple of high-$z$ DLAs 
\citep[\zabs = 2.087 towards Q~1444+0126 and \zabs = 2.426 towards Q~2348-0108 
discussed in ][]{Ledoux03, Noterdaeme07}. In comparison
to the two known \h2 systems at $z\le 1$, the  derived $f$(\h2) here is close to what
is measured in the \zabs = 0.56 sub-DLA towards Q~0107-0232 \citep{Crighton13} 
and an order of magnitude smaller than that measured in the 
\h2 system at \zabs = 0.18 towards B~0120-28 that also shows HD
molecules \citep{Oliveira14}. Equilibrium \h2 abundance is controlled by the  formation and the
destruction rates. In addition, as \h2 is optically thick in the present case, the effect 
of shielding becomes very important in establishing the equilibrium \h2 abundance. 

It is well known that the \h2 excitation temperature measured using $J=0$ and $J=1$ levels
(called T$_{01}$) traces the kinetic temperature of the gas very well when 
log~$N$(\h2)$\ge15.8$ \citep[see][]{Roy06}. Using Eq.~3 in \citet{Srianand05},
we estimate $86\le {\rm T_{01}} \le 270$ K,  $78\le {\rm T_{02}} \le 133$ K 
and  $95\le {\rm T_{13}} \le 135$ K.
%
This suggests that rotational level
populations are consitent with a single excitation temperature in the
range 95$-$133 K. 
%
This temperature range is slightly 
higher than that measured in the Galactic ISM \citep[77$\pm$17 K;][]{Savage77}
and Magellanic clouds \citep[82$\pm$21 K;][]{Tumlinson02} and consistent with
the lower end of what is measured in high-$z$ DLAs \citep[100$-$300 K;][]{Srianand05}.
Below we discuss the implications of equilibrium abundance of \h2 and the
rotational populations using photoionization code {\sc cloudy}.

\section{Photoionization models using Cloudy}
\begin{figure}
\centerline{
\includegraphics[width=8cm,bb=25 120 600 710,clip=,angle=0]{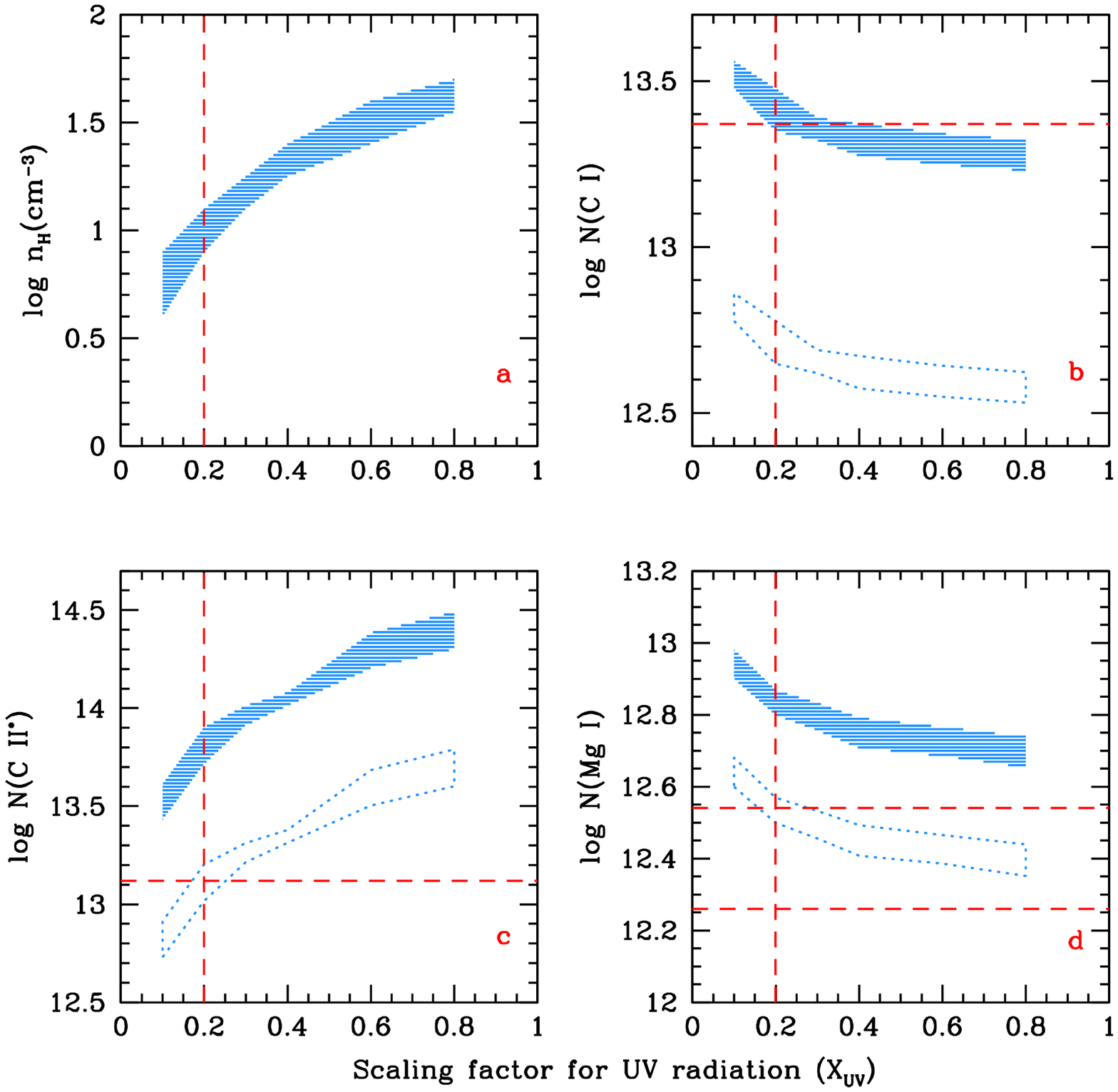} 
}
\caption{
Results of photoionization models using {\sc cloudy}. The shaded region in
panel (a) gives the allowed density range that reproduces the observed
$N$(\h2) for a given intensity of the background radiation field 
(denoted by $\chi_{UV}$). Shaded regions in the remaining panels
are the predicted range in the column density of a given species
for the density range given in panel (a). Regions marked by dotted
lines are the same after taking into account a typical elemental
depletion. Horizontal dashed lines are the measurements when
there are two lines and upper limit when only one line is present. 
}
\label{fig_mod}
\end{figure}

In this section we consider photoionization models constructed using
{\sc cloudy}. In order to get an accurate \h2 equilibrium
abundance, we use the full $J$ resolved calculations
as described  in \citep[][]{Shaw05}  using ``{\sc atom \h2}'' command.
Given the uncertainties associated with the H~{\sc i}
contribution to different phases we do not attempt to model all
the absorption species detected in the COS spectrum and focus mainly
on \h2, C~{\sc i}, Mg~{\sc i} and C~{\sc ii$^*$}. To start with
we assume that all the measured $N$(H~{\sc i}) is associated to the
\h2 component\footnote{However, it is worth remembering that at high-$z$ 
it is now recognized that only a small
fraction of H~{\sc i} may be associated with the \h2 component
\citep[see][]{Srianand12}.} and consider the absorbing gas to be a
single component. 

The ionizing radiation considered in our model is a combination of
meta-galactic UV background contributed by QSOs and galaxies 
\citep{Haardt96} and the interstellar radiation field similar to
that of the Milky Way\citep{Habing68} but scaled by $\chi_{UV}$. The value
of $\chi_{UV}$ equals to 0 and 1 corresponds to the UV radiation field  similar to 
the meta-galactic UV backgrond and that seen in the Milky Way disk
respectively.
The absorbing gas is
assumed to be a plane-parellel slab with the radiation field
illuminating it from one side\footnote{We note that considering two sided
illumination does not change our results as found by \citet{Dutta14}.}. The calculation is stopped
when $N$(H~{\sc i}) in the model is equal to the observed value.
We assume the metallicity and the dust depletion similar to what is observed
for this system and assumed the dust composition to be similar to 
Milky Way. For most of the models discussed here, we assume cosmic-ray
ionization rate, log($\Gamma_{CR}$) = $-$17.3, the default value in {\sc cloudy} taken from
\citet{Williams98}. Latter we will explore the effect of using different
values of $\Gamma_{CR}$.

\subsection{Constant temperature models}

First we run a set of constant temperature models keeping the
gas temperature to be 100 K as inferred from \h2 observations.
We find the allowed hydrogen density range ($n_H$) for a range
of $\chi_{UV}$. These are summarised in panel (a) in Fig.~\ref{fig_mod}.
In panel (b), we show the predicted values of $N$(C~{\sc i}) for this 
allowed density range (shaded region) for each value of  $\chi_{UV}$. The dashed
horizontal line is the observed 3$\sigma$ upper limit. It is clear
from this figure that  model predictions are consistent with
the observed $N$(C~{\sc i}) upper limits, when $\chi_{UV} \ge 0.2$. In this range the 
\h2 observations are consistent with $n_H\ge 10$ cm$^{-3}$.

It is clear from panel (d) in Fig.~\ref{fig_mod} that for the range in $\chi_{UV}$
and $n_H$ the predicted Mg~{\sc i} column densities are
0.2 dex higher than the observed value.  However, it is
well known that Mg is depleted by about 0.3 dex even in 
the halo gas of Milky way \citep[see Table 5, of][]{Welty99a}.
The dottes region marks the model prediction when Mg is
depleted by 0.3 dex. This is consistent with the observations.
Thus C~{\sc i}, Mg~{\sc i} and \h2 observations are consistently
reproduced by the models when a moderate Mg depletion is assumed
for $n_H\ge10$ cm$^{-3}$. We also notice that the model predicted
Cl~{\sc i} column density is also consistent with the measured
upper limits.

However, these models have problem reproducing the C~{\sc ii}$^*$
column density.
Panel (c) shows the predicted C~{\sc ii$^*$} column density as a function of 
$\chi_{UV}$ for the allowed $n_H$ range given in panel (a). 
It is clear (from the shaded region) that when the relative abundance
of C with respect to other elements is solar our models over 
predict C~{\sc ii$^*$} column density by more than 0.6 dex.
The deviation is more for higher values of $\chi_{UV}$.
The typical C depletion seen in our Galaxy (i.e., 0.4 dex)
alone can not bring the model predicted $N$(C~{\sc ii$^*$})
below the upper limit from observations. We need carbon
to be underabundant by more than 0.6 dex (see the region 
covered by the dotted line). Such underabundance of C is not
usually seen in DLAs but seen in halo stars with metallicity
similar to the present DLA \citep[see][]{Fabbian09}.
One possible way of avoiding the high C depletion is to
produce the required amounts of \h2 at lower densities.
This we shall explore below.


\subsection{Thermal equilibrium models}

We run next set of {\sc cloudy} simulations allowing the temperature of the gas
to be self-consistently computed by {\sc cloudy}. We consider
log($\Gamma_{CR}$)~=~$-17.3$ and $0.3\le \chi_{UV}\le 0.6$. We find
the observed \h2 column density is reproduced for $n_H$ in the
range 20 to 60 cm$^{-3}$. These models have log~$N$(C~{\sc i}),
and log~$N$(Mg~{\sc i}) in the range 13.69$-$13.78 and 13.03$-$13.10 
respectively. These are typically 0.3 dex higher than what we 
found for constant temperature model above.  We notice that
this is mainly due to the gas kinetic temperature in the
model being very low (i.e., T = 32 K). Clearly we need additional
heating for the gas to have correct \h2 excitation as well.
Such a situation is also encountered while modelling the
ISM sightlines using {\sc cloudy}\citep[see][]{Shaw06,Shaw08}.

Enhancement in the cosmic-ray ionization rate has been suggested as 
a possible solution to get high temperature in the models. 
To explore this, we consider a model with $\chi_{UV}\le 0.6$ and 
log($\Gamma_{CR}$)$= -15.7$.  The observed \h2 abundance is obtained
for $1.3\le log~n_H[cm^{-3}] \le 1.5$. This is similar to what 
we have in Fig~\ref{fig_mod}. As expected the gas temperature in enhanced
and is between 47-60 K. However, C~{\sc i} and Mg~{\sc i} column
densities predicted by these models are much higher than the observed
values. Thus we do not favor high $\Gamma_{CR}$ as the reason for 
high gas temperature.

Next, we considered the stoping column density $N$(H~{\sc i}) = 20.25 (a factor 2
smaller than the observed value) to mimic the case where
only part of the H~{\sc i} column density is associated
to the \h2 component or the gas disk is observed at an inclination angle 
\citep[see][]{Dutta14}. For simplicity, we considered $\chi_{UV} = 0.5$ and
varied log($\Gamma_{CR}$) in the range $-15.7$ to $-17.3$. We note that
observed constraints from \h2, C~{\sc i}, C~{\sc ii$^*$} and Mg~{\sc i}
can be satisfied when 1.6$\le log(n_H[cm^{-3}]) \le 1.8$, $log(\Gamma_{CR}) = -17.3$ 
and assume Mg and C are depleted by 0.3 dex and $\ge$ 0.3 dex respectively.
However, like the models discussed above the best fitted models have
kinetic temperature of 30 K. The models with high $\Gamma_{CR}$ gives 
slightly higher temperature albeit with increased abundances of Mg~{\sc i}
and C~{\sc i}.

Low temperatures produced in {\sc cloudy} models suggest that additional
non-radiative heating processes that are not included in {\sc cloudy} may be
playing a crucial role in heating the gas. Indeed numerical simulations
of SNe driven interstellar medium use energy from the SNe in various forms
to get the gas temperature  \citep[see for example,][]{Korpi99,Gazol01,Maclow05,Joung09}.
If such a mechanical injection of energy is available in the present system
also then the excess temperature noted can be explained.

\subsection{Models with enhanced \h2 formation rate:}

From the discussions presented above, it appears that the best solution to 
the lack of C~{\sc i} and C~{\sc ii$^*$} absorption and high kinetic temperature
is to produce the required amounts of \h2 at lower densities. One of the 
possibilities is to increase the \h2 formation rates on top of the dust grains.
Note, it is a general procedure to use the \h2 formation rate (i.e., $3\times10^{-17}$
cm$^{3}$ s$^{-1}$) inferred by
\citet{Jura74b} in the analytic calculations to model
systems with \h2 detections. However, there are indications that these
mean \h2 formation rate may not be sufficient to explain properties of
Photo-Dissociation Regions (PDR) with low to moderate excitation and enhanced formation rate may be 
needed \citep[see][]{Habart11, Lebourlot12}. To explore this,
we run {\sc cloudy} with an enhanced \h2 formation rate by a factor 2 to that
of \citet{Jura74b}. We note that the observered constraints can be
explained by a model with $\chi_{UV}\sim 0.2$, 0.6$\le log(n_H[cm^{-3}])\le$ 0.8.
However, these models still have gas kinetic temperature ($\sim 50 K$) a factor 
2 smaller than what we infer from the data. \citet{Lebourlot12} have suggested
that the grain physics included in their calculations that enhances the
\h2 formation rate can also produce formation heating. As these dust
processes are not yet incorporated in {\sc cloudy} we could not check
whether the additional formation heating can enhance the temperature to
the observed value.

\section{Galaxy candidates}

\begin{figure}
\centerline{
\includegraphics[height=7cm,width=10cm,angle=0]{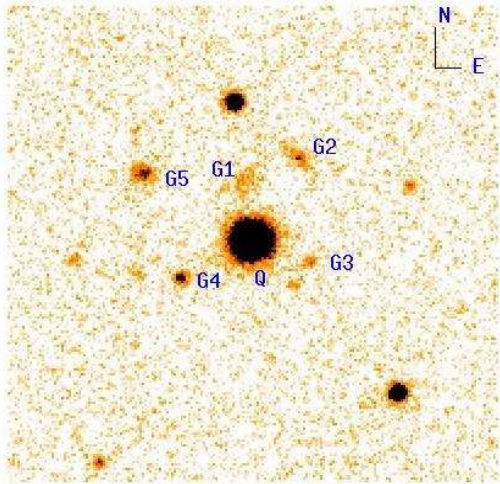} 
}
\caption{R-band image covering 1'$\times$1' centered around the QSO. The QSO and
5 galaxies close to the QSO in the angular scale are marked. Only galaxy
G1 has the photometric redshift consistent with the \zabs of the DLA. We 
consider this as the candidate host galaxy of the DLA. The seeing measured
in this image is close to 1.4 arc sec.}
\label{igo_r}
\end{figure}
 
We do not detect any line emission from the intervening galaxy in the SDSS fiber 
spectrum of the QSOs \citep[for example, 
as seen in][]{Noterdaeme10o3}. Using the measured signal-to-noise and a line width of 
300 \kms we find a 3$\sigma$ upper limit on H$\alpha$ luminosity to be $1.63\times 10^{34}$ watts.
The upper limit for the H-$\alpha$ flux is consistent with
a surface star formation rate of $\le 2.3\times 10^{-2}$M$_\odot$ yr$^{-1}$ kpc$^{-2}$ assuming
the absorbing galaxy  fills the SDSS fibre. This limit is not
stringent enough and allows for star formation similar to what is measured
in the case of low-$z$ 21-cm absorbers that show emission lines in the
SDSS fibre spectra \citep[see Table 3 of][]{Gupta13}. In these cases one
sees low surface brightness star forming galaxies whose disk is pierced
by the QSO sight line. Alternatively it is possible that the absorbing
gas is at a high impact parameter to the host galaxy as usually
identified in the past  for low-$z$ DLAs\citep[for e.g, ][]{Rao03}.

In our IGO R and V band images we identify 5 galaxies
within 30 arc sec to the QSOs as shown in Fig.~\ref{igo_r}.
Among them four galaxies (excluding G1) have photometric 
redshifts (based on our's and available SDSS photometry)
greater than 0.3. The galaxy G1 has consistent photometric
redshift to the absorber albeit with large errors (i.e., $z_p = 0.10\pm0.08$). In the absence
of spectroscopic data we consider this as a possible candidate DLA 
galaxy for further discussions. This galaxy is at a projected separation
of 8 arcsec from the QSO sight line. This corresponds to an impact
parameter of 14.2 kpc for the assumed cosmology (i.e., Flat universe with
$\Omega_\Lambda$ = 0.73 and $\Omega_m =0.27$ and H$_0$ = 71 km/s/Mpc). The R and V band
magnitudes of this galaxy are 21.50 and 20.85 mag respectively.
Thus the identified galaxy is a sub-L$_*$ galaxy.  The inferred impact parameter
is less than the median impact parameter for low-$z$ DLAs
found by \citet{Rao03} and similar to what is found 
for $z\sim0.1$ Mg~{\sc ii} absorber \citep{Kacprzak11}.

\section{Summary and discussions}

We report the detection of \h2 in a DLA at \zabs = 0.0963 toward J1619+3342.
The inferred molecular fraction ($-2.11\le$log $f$(\h2)$\le -1.85$) for the 
observed log~$N$(H~{\sc i}) = $20.55\pm0.10$  is much 
higher than what has been seen in the Milky Way and in Magellanic clouds but consistent with
what is measured in high lattitude gas in the Milky way halo \citep{Gillmon06}. 
In this case it was suggested that the enhanced \h2 formation rate in the
dense compressed gas may be the reason for the enhanced \h2 even at low $N$(H~{\sc i}).
This is also similar to what has been observed in the \zabs = 0.56 sub-DLA towards
Q~0107-0232 \citep{Crighton13}.
Using the rotational excitation of \h2 we infer the kinetic temperature of the
gas to be in the range 95 to 133 K. This is slighly higher than the typical
temperature measured in the Galactic ISM. But consistent with what is seen
in high-$z$ DLAs. In the absence of fine-structure lines of C~{\sc i} and C~{\sc ii}
stringent density measurements are not possible.

The observed metallicity is [S/H] = $-0.62\pm0.13$ similar to what is seen
in the SMC. The observed depeletion [Fe/S] = $-1.00\pm0.17$ is much less
than what is seen in the Milky Way. We find $N$(Ti~{\sc ii})/$N$ (Ca~{\sc ii})$\sim$0.3
consistent with what is measured in the Milky Way and larger than what is
measured in Magellanic clouds. We also find  no relative depletion between
Fe~{\sc ii} and Ti~{\sc ii}. This is consistent with the absence of strong
radiation field effects. In summary, the inferred metallicity is similar to
the mean value measured in SMC sight lines with the dust depletion intermediate
between lower Milky Way halo gas and LMC sight lines. If the strong correlation
seen between  the average line of sight density and Ti depletion is applicable
in the present case also then we can conclude that the average density in the
present DLA may be lower than that typically seen in the Milky Way disk.

Unlike most of the high-$z$ DLAs, we do not detect C~{\sc i} or C~{\sc ii$^*$}.
Based
on the $N$(C~{\sc ii$^*$}) upper limits we estimate the gas cooling rate,
log $l_c \le -27.03$ erg s$^{-1}$ per  hydrogen atom.  Therefore, this system
belongs to the ``low-cool'' population of DLAs defined by \citet{Wolfe08}.
If C is not depleted then this also implies the average density may be less
than what is typically seen in diffuse molecular gas in the Galactic ISM.

We consider wide range of photoionization model to understand the physical
conditions in the \h2 components. When we consider all H~{\sc i} column
density measured along the line of sight is associated with the \h2 components
we find constraints for C~{\sc i}, C~{\sc ii$^*$} and Mg~{\sc i} can 
be explained only when a C is depleted by more than 0.6 dex as seen in the
low metallicity star in the Milky Way.  These models require hydrogen
density in the range, 1.0$\le$log(${\rm n_H}$)$\le$1.5. 
Alternatively the observed constraints
can be explained by assuming only part of the observed $N$(H~{\sc i}) is associated
to the \h2 component or the \h2 formation rate on the dust grain is at least
a factor 2 higher than the typical value used in the Milky Way ISM models. 
In the first case the required hydrogen density is slightly highter than
that quoted above. In the second case the required density range is 
0.60$\le$log(${\rm n_H}$)$\le$0.8.
However, all the photoionization models produce gas
temperature typically a factor 2 less than the observed value. We conjucture
that additional heating like SNe heating seen in the hydrodynamical ISM
simulations may also be important in this system. 

We do not find any signatures of starformation activities along the line of
sight based on the lack of emission line detection in the QSO fiber spectra.
We identify a sub-L$_*$ galaxy at a impact parameter of 14.2 kpc as a possible
candidate for the DLA galaxy. This 
is consistent with the \h2 gas either associated with a very low luminosity 
galaxy along the line of sight or in the halo of the identified galaxy.
Detailed spectroscopic study of faint galaxies around this QSO is important to further understand
this \h2 system without speculating too much.

\section{acknowledgements}

We thank the IGO staffs for there help during observations.


%
\def\aj{AJ}%
\def\actaa{Acta Astron.}%
\def\araa{ARA\&A}%
\def\apj{ApJ}%
\def\apjl{ApJ}%
\def\apjs{ApJS}%
\def\ao{Appl.~Opt.}%
\def\apss{Ap\&SS}%
\def\aap{A\&A}%
\def\aapr{A\&A~Rev.}%
\def\aaps{A\&AS}%
\def\azh{AZh}%
\def\baas{BAAS}%
\def\bac{Bull. astr. Inst. Czechosl.}%
\def\caa{Chinese Astron. Astrophys.}%
\def\cjaa{Chinese J. Astron. Astrophys.}%
\def\icarus{Icarus}%
\def\jcap{J. Cosmology Astropart. Phys.}%
\def\jrasc{JRASC}%
\def\mnras{MNRAS}%
\def\memras{MmRAS}%
\def\na{New A}%
\def\nar{New A Rev.}%
\def\pasa{PASA}%
\def\pra{Phys.~Rev.~A}%
\def\prb{Phys.~Rev.~B}%
\def\prc{Phys.~Rev.~C}%
\def\prd{Phys.~Rev.~D}%
\def\pre{Phys.~Rev.~E}%
\def\prl{Phys.~Rev.~Lett.}%
\def\pasp{PASP}%
\def\pasj{PASJ}%
\def\qjras{QJRAS}%
\def\rmxaa{Rev. Mexicana Astron. Astrofis.}%
\def\skytel{S\&T}%
\def\solphys{Sol.~Phys.}%
\def\sovast{Soviet~Ast.}%
\def\ssr{Space~Sci.~Rev.}%
\def\zap{ZAp}%
\def\nat{Nature}%
\def\iaucirc{IAU~Circ.}%
\def\aplett{Astrophys.~Lett.}%
\def\apspr{Astrophys.~Space~Phys.~Res.}%
\def\bain{Bull.~Astron.~Inst.~Netherlands}%
\def\fcp{Fund.~Cosmic~Phys.}%
\def\gca{Geochim.~Cosmochim.~Acta}%
\def\grl{Geophys.~Res.~Lett.}%
\def\jcp{J.~Chem.~Phys.}%
\def\jgr{J.~Geophys.~Res.}%
\def\jqsrt{J.~Quant.~Spec.~Radiat.~Transf.}%
\def\memsai{Mem.~Soc.~Astron.~Italiana}%
\def\nphysa{Nucl.~Phys.~A}%
\def\physrep{Phys.~Rep.}%
\def\physscr{Phys.~Scr}%
\def\planss{Planet.~Space~Sci.}%
\def\procspie{Proc.~SPIE}%
\let\astap=\aap
\let\apjlett=\apjl
\let\apjsupp=\apjs
\let\applopt=\ao
\bibliographystyle{mn}
\bibliography{mybib}
\end{document}